\documentclass{jetp}
\twocolumn

\usepackage{hyperref}

\newcommand{\eq}[2]{\begin{equation} \label{e:#1} #2 \end{equation}}
\newcommand{\eql}[2]{\begin{multline} \label{e:#1} #2 \end{multline}}

\newcommand{\re}[1]{(\ref{e:#1})}

\newcommand{\lrb}{\left(}
\newcommand{\rrb}{\right)}
\newcommand{\ltb}{\left<}
\newcommand{\rtb}{\right>}

\newcommand{\dl}[1]{{\Delta_{#1}}}

\lat

\title{Non-conformal limit of AGT relation \\ from the 1-point torus conformal block}

\rtitle{Non-conformal limit of AGT relation from the 1-point torus conformal block}

\sodtitle{Non-conformal limit of AGT relation for the 1-point torus conformal block}

\author{Vasiliy Alba $^{\star\dag\ddag\flat}$,
Andrey Morozov$^{\star\intercal}$\/\thanks{e-mail:alba@itp.ac.ru, e-mail:andrey.morozov@itep.ru}}

\rauthor{V.Alba, And.Morozov}

\sodauthor{Alba, Morozov}

\address{$^{\star}$ITEP, Moscow 117218, Russia\\
$^\dag$Landau Institute for Theoretical Physics,
119334 Moscow, Russia\\
$^\ddagger$Department of General and Applied Physics, MIPT, Moscow 117303, Russia\\
$^\intercal$Physical Department, MSU, Moscow 119991, Russia\\
$^\flat$ Bogolyubov Institute for Theoretical Physics, Kyiv 03680, Ukraine}

\abstract{
Given a $4d$ $\mathcal{N}=2$ SUSY gauge theory, one can construct the Seiberg-Witten prepotentional,
which involves a sum over instantons. Integrals over instanton moduli spaces require regularisation.
For UV-finite theories the AGT conjecture favours particular, Nekrasov's way of regularization.
It implies that Nekrasov's partition function equals
conformal blocks in 2d theories with $W_{N_c}$ chiral algebra.
For $N_c=2$ and one adjoint multiplet it coincides with a torus 1-point Virasoro conformal block.
We check the AGT relation between conformal dimension and adjoint multiplet's mass
in this case and investigate the limit of the conformal block, which corresponds to the large mass
limit of the $4d$ theory e.i. the asymptotically free $4d$ $\mathcal{N}=2$ supersymmetric Yang-Mills theory.
Though technically more involved, the limit is the same as in the case of fundamental
multiplets, and this provides one more non-trivial check of AGT conjecture.
}


\begin{document}

\maketitle

\section{Introduction}

$\mathcal{N}=2$ supersymmetric Yang-Mills (SYM) theories have attracted attention for rather a long time, because they are ideally suited for the study of interplay between perturbative and
non-perturbative effects and for manifestation of various dualities \cite{n21}-\cite{n23}.
Depending on the fields content, these theories exhibit all types of renormalization behaviour of effective coupling constant $g$: it may tend to infinity (Landau p\^{o}le),
and to zero (asymptotic freedom with dimensional transmutation in IR) or remain constant (UV-finite).

In $\mathcal{N} = 2$ SYM theory the low-energy effective action is Abelian and its most important part is expressed in terms of the prepotential.
Prepotential contains one-loop perturbative contribution and a far more sophisticated
non-peturbative part, obtained as a sum over instantons. It was explicitly found by N.Seiberg and E.Witten (SW) \cite{n21, n24} with the help of duality arguments, and the answer was soon
reformulated in terms of the spectral surfaces and simple integrable systems \cite{gk3m, dw}.
The spectral curves were later interpreted in terms of branes.
Straightforward evaluation of instanton sums
is rather difficult, especially because
some of the integrals over instanton moduli
spaces diverge.
See \cite{khose} for a comprehensive
review and references.

A very successful direct caluculation was finally
provided by N.Nekrasov \cite{nek}.
He introduced a new partition function,
depending on additional parameters $\epsilon_1$
and $\epsilon_2$, such that the limit $\epsilon_1, \epsilon_2\longrightarrow 0$ reproduces SW prepotential.

Recently F.Alday, D.Gaiotto and Y.Tachikawa (AGT) made a ground-breaking conjecture
that Nekrasov functions coincide with conformal
blocks \cite{bpz} of $2d$ Liouville/Toda models, and
the $\epsilon$-parameters are needed to allow
arbitrary values of the central charge in their
chiral $W_{N_c}$ algebras (for $N_c=2$ the chiral
algebra is just the ordinary Virasoro).
AGT suggest a non-trivial association of conformal
blocks with UV-finite $4d$ quiver models.
The $4$-point tree Virasoro block is associated
with the $N_c=2$ gauge theory with $2N_c=4$
additional fundamental matter supermultiplets.

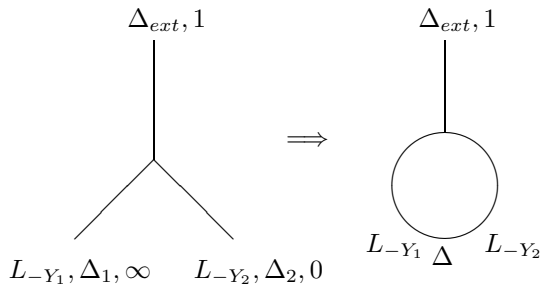
\begin{figure}[h]
\begin{picture}(70, 120)(-20, 0)
\label{fig:ipt}
\put(170, 65){\line(0, 1){35}}
\put(160, 105){$\Delta_{ext}, 1$}
\put(170, 45){\circle{40}}
\put(140, 20){$L_{-Y_1}$}
\put(185, 20){$L_{-Y_2}$}
\put(165, 15){$ \Delta$}

\put(90, 25){\line(-1, 1){30}}
\put(30, 25){\line(1, 1){30}}
\put(60, 55){\line(0, 1){45}}
\put(50, 105){$\Delta_{ext}, 1$}
\put(5, 10){$L_{-Y_1}, \Delta_1, \infty$}
\put(75, 10){$L_{-Y_2}, \Delta_2, 0$}

\put(110, 60){$\Longrightarrow$}
\end{picture}
\caption{\footnotesize{Triple vertex with two Virasoro descendants and the 1-point
toric conformal block, obtained by taking a trace over
Vermat module with a given dimension $\Delta$.
Each line is charaterized by dimension, by Ferrers diagram
and external legs are also labeled by the position of
the vertex operator on the Riemann surface.}}
\end{figure}
If there is instead, a single adjoint matter multiplet
which also makes $4d$ theory UV-finite,
the associated conformal block is the toric 1-point
function. This claim was made in \cite{agt} and partly
checked in \cite{pog}. We also confirm this relation
and check it in one more way. Namely, we consider
the limit of the large mass of adjoint multiplet,
where it decouples and the $4d$ theory turns into
asymptotically free pure gauge $N=2$ SYM.
This pure gauge theory can be also obtained
as the large-mass limit of the theory with $4$
fundamentals, which implies that the corresponding
limits of the tree $4$-point and the toric $1$-point
conformal blocks should be the same.
The first limit has already been studied in \cite{3mnc, gai}.
We find the second limit \re{fiex} and show that it is indeed
the same.

\section{AGT relations}
 AGT hypothesis consists of several statements about relations between $2d$ CFT and $4d$ $\mathcal{N}=2$ SYM theories.
One of the statements is that perturbative part of Nekrasov partition function is equal to the product of DOZZ factors \cite{ZZ, do},
defining dependence of the triple functions in $2d$ Liouville theory on dimensions.
Even more important and interesting is
another part of this conjecture:
the instanton part of Nekrasov partition function
is equal to conformal block in $2d$ CFT
(which depends on the chiral algebra, but not on
the other details of $2d$ conformal model).
Many examples were considered in \cite{agt} and later
discussed in some detail \cite{rp1}-\cite{rp18}, \cite{pog}

A list of many Nekrasov functions is available
in numerous papers, starting from original \cite{nek}.
More difficult is the situation with conformal blocks.
Like Nekrasov functions they are formal series;
in the simplest cases of interest in the present
paper they are in one variable,
\eq{cb}
{
\mathcal{B}(x)=\sum\limits_{n=0}^{\infty}
x^n\mathcal{B}^{(n)},
}
$n$ is called the "level", and particular quantities
$\mathcal{B}^{(n)}$ are built from two kinds of ingredients:
Shapovalov form
\eq{shf}
{
Q_{\Delta}(Y_1, Y_2)=\frac{\left<L_{-Y_1}V_{\Delta}\Big|L_{-Y_2}V_{\Delta}\right>}
{\left<V_{\Delta}\Big|V_{\Delta}\right>}
}
and two kinds of triple vertices \cite{4m}
\eq{cor}
{
\gamma_{123}(Y_1, Y_2, Y_3)=
\frac{\left<L_{-Y_1}V_1(0) L_{-Y_2}V_2(1) L_{-Y_3}V_3(\infty)\right>}
{\left<V_1(0) V_2(1) V_3(\infty)\right>}
}
and
\eq{scpr}
{
\bar\gamma_{12;3}(Y_1, Y_2, Y_3)=
\frac{\left<L_{-Y_3}V_3\Big| L_{-Y_1}V_1(1) L_{-Y_2}V_2(0)\right>}
{\left<V_3\Big| V_1(1) V_2(0)\right>}.
}
Here V are vertex operators, satisfying
operator product expansions
\eq{2to1}
{
 V_1(x_1)V_2(x_2)=\sum\limits_k (x_1-x_2)^{\Delta_1+\Delta_2-\Delta_k} C_{12}^k V_k(x)
}
Operators are made from primaries by the
action of Virasoro generators. Virasoro descendants
are labeled by Young-Ferrers diagrams $Y_i$.
Ferrers diagram is a sequence of integer numbers $k_1\geq k_2\geq k_3...$.
So we define $L_{-Y}$ as $L_{-Y}V=...L_{-k_3}L_{-k_2}L_{-k_1}V$.

Using the integral definition of Virasoro operators one can get the following relation:
\eql{Vflip}
{
\ltb L_{-n} V_1\Big| V_2(1)V_3(0)\rtb=
\ltb V_1\Big|V_2(1)(L_{n}V_3)(0)\rtb+\\
+\ltb V_1\Big|(L_{-1} V_2)(1)V_3(0)\rtb+(1+n)\dl{2} \ltb V_1\Big|V_2(1)V_3(0)\rtb+\\
+\sum_{k>0}C_{1+n}^{k+1}\ltb V_1\Big| (L_k V_2)(1)V_3(0)\rtb, \ \forall n.
}
It is valid for arbitrary fields $V_i$, not obligatory primary ones \cite{4m}.
Using this formula we can calculate all needed $\bar\gamma_{12;3}$.
The 4-point conformal block was computed already by many authors,
because in this case we need only $\bar\gamma_{12;3}(\emptyset, \emptyset, Y)$,
for which there is the well known general formula. The 1-point torus conformal block, which is of interest for us here, is made from a more complicated
$\bar\gamma_{12;3}(\emptyset, Y_2, Y_3)$, which is not yet known in the
general form. Thus we need to compute these vertices one by one.

Writing the correlator of 4 fields and expanding it with the help of \re{2to1} and using recently introduced notations we get
\eq{4cb}
{
\mathcal{B}^{(n)}_{4-point}=\sum\limits_{|Y_{\alpha}|=|Y_{\beta}|=n}\bar{\gamma}_{12;\alpha}(Y_{\alpha})
Q^{-1}_\Delta(Y_{\alpha}, Y_{\beta}) \gamma_{\beta 34}(Y_{\beta}).
}
It is clear that to compute the conformal block one should use
$\gamma$ instead of each vertex and $Q^{-1}$ instead of inner lines.

From the AGT relation for the 4-point conformal block we obtain
\eq{d1}
{
\Delta=\frac{\epsilon^2-4a^2}{4\epsilon_1\epsilon_2}, \ c=1+\frac{6\epsilon^2}{\epsilon_1\epsilon_2},
}
In these relations $a$ is a v.e.v. of the $4d$ SYM theory.
They were originally obtained in \cite{agt} and \cite{3m}.
Nevertheless we defined one external dimension and power of $\eta$-multiplier.

\section{1-point conformal block on a torus}
The formula to calculate 1-point torus conformal block is

\begin{align}
&\mathcal{B}(q)=\sum\limits_{n}q^n \mathcal{B}^{(n)}=\label{1pcb}\\
&=\sum\limits_{Y_1, Y_3} q^{|Y_1|}\left< L_{-Y_1}V_1\big| V_{ext}(1)L_{-Y_2}V_2(0)\right> Q^{-1}_{\Delta}(Y_1, Y_2)\notag
\end{align}
Besides $x$ it depends on two dimensions,
$\Delta$ and $\Delta_{ext}$ and on the central charge $c$.
AGT conjecture identifies this conformal block with
analogous expansion of Nekrasov partition function
\eq{nk}
{
\widetilde{\mathcal{N}}(q)=\lrb q^{-1/24}\eta(q)\rrb^{-\nu}\sum\limits_{n=0}^{\infty}
q^n\mathcal{N}^{(n)}=\sum\limits_{n=0}^{\infty}\widetilde{\mathcal{N}}^{(n)},
} where $\eta(q)=q^{\frac{1}{24}} \prod\limits_{n=1}^{\infty} (1-q^{n})$ is the Dedekind eta function, $q = e^{2\pi {\rm{i}} \tau}$, $\tau=\frac{4\pi i}{g^2}+\frac{\theta}{2\pi}$ is complex coupling constant. ${\cal N}^{(n)}$ depends on the v.e.v. modulus $a$,
on adjoint multiplet's mass $m$ and also on $\epsilon_1$
and $\epsilon_2$.

\subsection{The First Level}
The AGT relation $\{\Delta, \Delta_{ext}, c\}
\ \stackrel{\nu}{\leftrightarrow}\
\{a, m, \epsilon_1, \epsilon_2\}$
can be found from equality ${\cal B}^{(1)} = \widetilde{\mathcal{N}}^{(1)}$
at level one.
Explicitly
\eq{cb1lp}
{
\mathcal{B}^{(1)}=\frac{\Delta_{ext}^2}{2\Delta}-\frac{\dl{ext}}{2\Delta}+1,
}
while
\eql{n1l}
{
\mathcal{N}^{(1)}=
\frac{(\epsilon_1-m)(\epsilon_2-m)}{\epsilon_1\epsilon_2(\epsilon^2-4a^2)}(-8a^2+2\epsilon^2-2\epsilon m +2m^2).
}
These quantities coincide provided (1) is suplemented by
\eq{d2}
{
\dl{ext}=\frac{m(\epsilon-m)}{\epsilon_1\epsilon_2}, \
\nu=1-\frac{2m(\epsilon-m)}{\epsilon_1\epsilon_2}.
}

The answer was computed with the help of \emph{ad hoc} triple conformal correlator with a non primary field \cite{toapp}.
As we already noticed this computation is non trivial because it involves the vertex
$\bar\gamma_{23;1}$ with two non-trivial
Young diagrams, see \cite{toapp} for details.

\subsection{The Second Level}
The first non-trivial check of AGT conjecture is at level two.
We made this check and there is indeed a
complete coincidence between conformal block and
Nekrasov partition function holds at level two,
as already claimed in \cite{pog}.
Unfortunately, the full formula is too cumbersome to be
presented here.

Instead in this paper we concentrate on additional check,
which can be extended to all levels:
we investigate the limit of large $m$.
According to (13) this is the same as large $\Delta_{ext}$,
and what we need is a new asymptotics:
 (15)
Together with (14) this gives an insight: only
particular terms dominate in the limit.

\section{Large Mass Behaviour}

AGT relation is originally formulated for UV-finite gauge theories in $4d$.
Asymptotically free pure gauge theory arises when masses
of additional matter supermultiplets are led to infinity,
while the bare coupling $x \sim q_0$ is simultaneously
led to zero.
In the case of adjoint multiplet the product
$xm^2 = \Lambda^4$ is kept constant in this scaling limit.
We know that if we have a large mass, we also have large $\dl{ext}$.

With the aid of \re{d2} one can obtain asymptotic behaviour of the first and the second order term of the conformal block
\eq{cb1l}
{
\mathcal{B}^{(1)} \mathop{\asymp}_{m\longrightarrow \infty} \frac{\Delta_{ext}^2}{2\Delta}=
\Delta_{ext}^2 Q_{\Delta}^{-1}\lrb [1], [1]\rrb.
}
\eq{cb2l}
{
\mathcal{B}^{(2)}\mathop{\asymp}_{m\longrightarrow \infty}
\Delta_{ext}^4 Q_{\Delta}^{-1}\lrb [1^2], [1^2]\rrb.
}

Substituting $\Lambda$ in \re{cb1l}, \re{cb2l} and \re{cb} and generalizing this formula
we can guess that the large mass limit of $\mathcal{B}$ looks like
\eq{fiex}
{
\boxed{
\mathop{\lim\limits_{m\longrightarrow \infty}}_{xm^2=\Lambda^4=const}\mathcal{B}(x)=\sum\limits_{n=0}^{\infty}
\Lambda^{4n} Q_{\Delta}^{-1}\lrb [1^n], [1^n]\rrb
}}

We also have an explanation why this formula is correct.
When one is studying high mass or in other notations large $\dl{ext}$ limit,
one should focuse only at the term with the highest power of $\dl{ext}$, because of \re{d2}.

First of all we should describe how we evaluate $\mathcal{B}^{(n)}$:

\eq{Bfrg}
{
\mathcal{B}^{(n)}=\sum\limits_{|Y_i|= |Y_j|=n}\ltb L_{-Y_i} V_1\Big| V_2(1)L_{-Y_j}V_3(0)\rtb Q_{\Delta}^{-1}(Y_i, Y_j)
}
One can see that $Q$ depends only on $\Delta$, so all $\dl{ext}$ dependence is concentrated in $\gamma$.
We will prove a theorem that the highest power of $\dl{ext}$ in
\eq{gamma}{
\ltb L_{-Y_i} V_1\Big|V_2(1)L_{-Y_j}V_3(0)\rtb
}
is equal to the total number of Virasoro operators in $Y_i$ and $Y_j$.
If this theorem is correct then the highest power of $\dl{ext}$ is in the term where $Y_i=Y_j=[1^n]$.
We also will prove that the coefficient in front of the highest power of $\dl{ext}$ in the scalar product
for this term is equal to one.
Therefore the leading term in $\mathcal{B}^{(n)}$ when $\dl{ext}$ is large indeed looks like:
$
\Delta^{2n}Q_{\Delta}^{-1}\lrb [1^n], [1^n]\rrb
$

\section{Proof of the theorem}

In this section we prove that the highest order $\Delta_{ext}$ in scalar product
of three fields \re{gamma} is equal to the total number of Virasoro operators in this scalar product.
Also the coefficient in front of this highest order term will be found.

Using relation \re{Vflip} with $n=0$ one can show that
\eq{O1O}
{
\ltb V_1\Big| L_{-1}V_2(1) V_3(0)\rtb=\widetilde{\Delta}_{ext}\ltb V_1\Big| V_2(1) V_3(0)\rtb.
}
Here and below we use the notation
$\widetilde{\Delta}_{ext}=\dl{ext}+\left(\text{\small{arbitrary\ function\ of\ $\Delta$}}\right)$ and
we can use $\widetilde{\Delta}_{ext}$ instead of $\dl{ext}$ because we are only interested in the highest power of the $\dl{ext}$ term.

We can use the induction method to prove our statement.
So the base of induction is the zero level -
if there are no Virasoro operators then $\gamma$ does not depend on $\dl{ext}$ because it is equal to one.
Using \re{O1O} we can see that two terms in the second row of \re{Vflip} can be summed directly and give us $n\widetilde{\Delta}_{ext}$ along with
lowering the number of the Virasoro operators
in front of $V_1$ by one.
From the Virasoro commutation relation we know that
$L_k L_{-n}V_3=L_{-n} L_k V_3+(n+k)L_{k-n} V_3+\delta_{k, n}\frac{cm(m^2-1)}{12}$.
So if $V_3$ is a primary field, as it is in our case,
$L_k L_{-Y} V_3$ is the sum of the fields with the same number of operators as $L_{-Y} V_3$,
and the coefficients of this sum depends only on $\Delta$.
So the order of this term is lower than the order of the second row in \re{Vflip}.
The third row in our case is equal to zero, because in our case $V_2$ is always a primary field.

So using previous statements one can see
that reducing the number of the operators in front of $V_1$ gives us
multiplication by $n\widetilde{\Delta}_{ext}$ from the second row plus some terms of the lower order,
where $n$ is the order of the Virasoro operator which we expanded.
And similarly reduction of the number of operators in front of $V_3$ also give the $n\widetilde{\Delta}_{ext}$ multiplier.

Therefore each time when one reduces the total number of operators in \re{gamma} by one he gains a product of $\widetilde{\Delta}_{ext}$, the function of $\Delta$ and the expression of similar form but lower order.
Also one can see that the coefficient in front of the highest power of $\dl{ext}$ in \re{gamma} is equal to
\eq{}{
\prod\limits_{L_{-p}\in Y_i}p\prod\limits_{L_{-q}\in Y_j}q
}
So we not only proved that the highest power is equal to the total number of Virasoro operators but also counted the coefficient in front of this term.
And indeed this coefficient is equal to one if $Y_i=Y_j=[1^n]$.

\section{Conclusion}
We studied the large-mass limit of the AGT relation
for the $N=2$ SYM theory with $N_c=2$ and adjoint matter multiplet.
The corresponding limit of the 1-point conformal block
on a torus reproduces the answer, obtained earlier
by taking the similar limit of the 4-point tree
conformal block, associated through AGT relation
with the $4d$ theory with $N_f=2, N_c=4$ fundamental
multiplets.
The fact that the two limits coincide is implied
by AGT relation, but is somewhat non-trivial from
the point of view of CFT.
Identity involves two very different conformal blocks,
and even the relevant triple vertices are different
in two cases: in toric case more sophisticated
vertices with two Virasoro descendants are needed.
Thus we obtained one more non-trivial confirmation of
the AGT conjecture.

\section*{Acknowledgements}
We appeciate very useful discussions with Andrei Mironov and Alexei Morozov.
Also we are grateful to Alexander Belavin for his excellent lectures on CFT.

Vasiliy Alba's research was held within the bounds of Federal Program "Scientific and Scientific-Pedagogical personnel of innovational Russia" on 2009-2013 y.,Goskontrakt N P1339 and was partly supported by RFBR grant 10-02-00499%, NSh-3035.2008.2
. The work of Andrey Morozov is partly supported by
the grants RFBR 07-01-00526 and NSh-3036.2008.2.


\begin{thebibliography}{99}

\bibitem{n21} N.Seiberg and E.Witten,
 \emph{``Electric-magnetic duality, monopole condensation, and confinement in $\mathcal{N} = 2$ supersymmetric Yang-Mills theory''}, Nucl. Phys. {\bf B 426}, 19 (1994),
 \texttt{ arXiv:hep-th/9407087}.

\bibitem{n24} N.Seiberg, E.Witten,
 \emph{``Monopoles, duality and chiral symmetry breaking in $\mathcal{N}=2$ supersymmetric QCD''}, Nucl.Phys. {\bf B 431}, 484-550 (1994),
 \texttt{ arXiv: hep-th/9408099}

\bibitem{n22} Ph.C.Argyres and A.E.Faraggi,
 \emph{`` The vacuum structure and spectrum of $\mathcal{N} = 2$ supersymmetric $SU(N)$ gauge theory''}, Phys.Rev.Lett. {\bf 74}, 3931-3934 (1995),
 \texttt{ arXiv:hep-th/9411057}.

\bibitem{n23} A.Hanany and Y.Oz,
 \emph{``On the quantum moduli space of vacua of $\mathcal{N} = 2$ supersymmetric $SU(N_c)$ gauge theories''}, Nucl.Phys. {\bf B 452} (1995), 283-312.
 \texttt{ arXiv:hep-th/9505075}.

\bibitem{gk3m} A.Gorsky, I.Krichever, A.Marshakov, A.Mironov, A.Morozov,
 \emph{``Integrability and Seiberg-Witten Exact Solution''}, Phys.Lett. {\bf B 355}, 466-474 (1995),
 \texttt{ arXiv:hep-th/9505035}.

\bibitem{dw} R.Donagi, E.Witten,
 \emph{``Supersymmetric Yang-Mills Systems And Integrable Systems''}, Nucl.Phys. {\bf B 460}, 299-334 (1996),
 \texttt{ arXiv:hep-th/9510101}.

\bibitem{khose} N.Dorey, T.J.Hollowood, V.V.Khoze, M.P.Mattis
 \emph{``The Calculus of Many Instantons''}, Phys.Rept. {\bf 371}, 231-459 (2002),
 \texttt{ arXiv:hep-th/0206063}.

\bibitem{nek} N.A.Nekrasov
 \emph{ ''Seiberg-Witten Prepotential from Instanton Counting''}, Adv. Theor. Math. Phys. Volume {\bf 7}, Number {\bf 5}, 831-864 (2003),
 \texttt{ arXiv:hep-th/0206161}.

\bibitem{bpz} A.A.Belavin, A.M.Polyakov, A.B.Zamolodchikov,
 \emph{``Infinite conformal symmetry in two-dimensional quantum field theory''}, Nucl. Phys. {\bf B 241}, 333-380 (1984).

\bibitem{agt} L.Alday, D.Gaiotto, Y.Tachikawa,
 \emph{ ``Liouville Correlation Functions from Four-dimensional Gauge Theories''},
 \texttt{ arXiv:0906.3219}.

\bibitem{pog} R.Poghossian,
 \emph{``Recursion relations in CFT and $\mathcal{N}=2$ SYM theory''},
 \texttt{ arXiv:0909.3412}.

\bibitem{3mnc} A.Marshakov, A.Mironov, A.Morozov,
 \emph{``On non-conformal limit of the AGT relations''},
 \texttt{ arXiv:0909.2052}.

\bibitem{gai} D.Gaiotto,
 \emph{``Asymptotically free $\mathcal{N}=2$ theories and irregular conformal blocks''},
 \texttt{ arXiv:0908.0307}.

\bibitem{ZZ} A.B.Zamolodchikov, Al.B.Zamolodchikov,
 \emph{``Structure Constants and Conformal Bootstrap in Liouville Field Theory''},
 Nucl.Phys. {\bf B 477}, 577-605 (1996),
 \texttt{ arXiv:hep-th/9506136}.

\bibitem{do} H.Dorn, H.-J.Otto,
 \emph{``Two and three-point functions in Liouville theory''}, Nucl.Phys. {\bf B 429}, 375-388 (1994),
 \texttt{ arXiv:hep-th/9403141}.

\bibitem{rp1} N.Wyllard,
 \emph{``$A_{N-1}$ conformal Toda field theory correlation functions from conformal $\mathcal{N}=2$ $SU(N)$ quiver gauge theories''},
 \texttt{ arXiv:0907.2189}.

\bibitem{rp2} N.Drukker, D.Morrison and T.Okuda,
 \emph{``Loop operators and S-duality from curves on Riemann surfaces''}, JHEP {\bf 0909}, 031 (2009),
 \texttt{ arXiv:0907.2593}.

\bibitem{3m} A.Marshakov, A.Mironov and A.Morozov,
 \emph{``On Combinatorial Expansions of Conformal Blocks''}
 \texttt{ arXiv:0907.3946}.

\bibitem{4m} A.Mironov, S.Mironov, A.Morozov and An.Morozov,
 \emph{``CFT exercises for the needs of AGT''},
 \texttt{ arXiv:0908.2064}.

\bibitem{rp6} A.Mironov and A.Morozov, Phys.Lett. {\bf B680}, 188-194 (2009),
 \emph{``The Power of Nekrasov Functions''},
 \texttt{ arXiv:0908.2190}.

\bibitem{rp7} A.Mironov and A.Morozov,
 \emph{``On AGT relation in the case of $U(3)$''}, Nucl.Phys. {\bf B 825}, 1-37 (2009),
 \texttt{ arXiv:0908.2569}.

\bibitem{rp8} S.Iguri and C.Nunez,
 \emph{``Coulomb integrals and conformal blocks in the $AdS_3-WZNW$ model''},
 \texttt{ arXiv:0908.3460}.

\bibitem{rp9} D.Nanopoulos and D.Xie,
 \emph{``On Crossing Symmmetry and Modular Invariance in Conformal Field Theory and S Duality in Gauge Theory''},
 \texttt{ arXiv:0908.4409}.

\bibitem{rp10} L.Alday, D.Gaiotto, S.Gukov, Y.Tachikawa and H.Verlinde,
 \emph{``Loop and surface operators in $\mathcal{N}=2$ gauge theory and Liouville modular geometry''},
 \texttt{ arXiv:0909.0945}.

\bibitem{rp11} N.Drukker, J.Gomis, T.Okuda and J.Teschner,
 \emph{``Gauge Theory Loop Operators and Liouville Theory''},
 \texttt{ arXiv:0909.1105}.

\bibitem{rp13} R.Dijkgraaf and C.Vafa,
 \emph{``Toda Theories, Matrix Models, Topological Strings, and $\mathcal{N}=2$ Gauge Systems''},
 \texttt{ arXiv:0909.2453}.

\bibitem{rp14} A.Marshakov, A.Mironov and A.Morozov,
 \emph{``Zamolodchikov asymptotic formula and instanton expansion in $\mathcal{N}=2$ SUSY $N_f=2N_c$ QCD''},
 \texttt{ arXiv:0909.3338}.

\bibitem{rp15} A.Mironov and A.Morozov,
 \emph{``Proving AGT relations in the large-c limit''},
 \texttt{ arXiv:0909.3531}.

\bibitem{rp16} G.Bonelli, A.Tanzini,
 \emph{``Hitchin systems, $\mathcal{N}=2$ gauge theories and $W$-gravity''},
  \texttt{ arXiv:0909.4031}	

\bibitem{rp17} A.Gadde, E.Pomoni, L.Rastelli and S.Razamat,
 \emph{``S-duality and $2d$ Topological QFT''},
 \texttt{ arXiv:0910.2225}.

\bibitem{rp18} L.Alday, F.Benini and Y.Tachikawa,
 \emph{``Liouville/Toda central charges from $M5$-brane''},
 \texttt{ arXiv:0909.4776}.

\bibitem{rp19} H.Awata and Y.Yamada,
 \emph{``Five-dimensional AGT Conjecture and the Deformed Virasoro Algebra''},
 \texttt{ arXiv:0910.4431}.

\bibitem{mmbz} A.Mironov, A.Morozov,
 \emph{``Nekrasov Functions and Exact Bohr-Sommerfeld Integrals''},
 \texttt{ arXiv:0910.5670}.

\bibitem{toapp} V.Alba, An.Morozov,
 \emph{``Check of AGT Relation for Conformal Blocks on Sphere''},
{\tt arXiv:0912.2535}

\end{thebibliography}
\end{document}